\documentclass[nofootinbib,aps,pra,showpacs,twocolumn,groupedaddress]{revtex4}
\usepackage{mathtools}
\usepackage{amssymb}
\usepackage{graphics}
\usepackage[dvips]{graphicx}
\usepackage{graphicx}
\usepackage{color}

\newcommand \beq{\begin{eqnarray}}
\newcommand \eeq{\end{eqnarray}}
\newcommand \be{\begin{equation}}
\newcommand \ee{\end{equation}}

\def\simge{\mathrel{%
       \rlap{\raise 0.511ex \hbox{$>$}}{\lower 0.511ex \hbox{$\sim$}}}}
\def\simle{\mathrel{
       \rlap{\raise 0.511ex \hbox{$<$}}{\lower 0.511ex \hbox{$\sim$}}}}

\newcommand{\mpl}{M_{\rm pl}}
\newcommand{\redflag}[1]{{\color{red} #1}}
\newcommand{\blueflag}[1]{{\color{blue} #1}}

\usepackage[colorlinks=true,linktocpage=true,linkcolor=blue,citecolor=blue]{hyperref}

\newcommand{\eq}[2]{\begin{equation}\label{#1}#2 \end{equation}}
\def\nn{\nonumber}

\begin{document}
\title{Damping of gravitational waves by matter}
\author{Gordon Baym,$^{a,b}$ Subodh P.\ Patil,$^{b}$ and C.\ J.\ Pethick$^{b,c}$}
\affiliation{\mbox{$^a$Department of Physics, University of Illinois, 1110
  W. Green Street, Urbana, IL 61801-3080} \\
\mbox{$^b$The Niels Bohr International Academy, The Niels Bohr Institute,}\\\mbox{ University of Copenhagen, Blegdamsvej 17, DK-2100 Copenhagen \O,
 Denmark}\\
\mbox{$^c$NORDITA, KTH Royal Institute of Technology and Stockholm University,}\\
\mbox{
Roslagstullsbacken 23, SE-10691 Stockholm, Sweden }\\
}

\date{\today}

\begin{abstract}

We develop a unified description, via the Boltzmann equation, of damping of gravitational waves by matter, incorporating collisions. We identify two physically distinct damping mechanisms -- collisional and Landau damping.  We first consider damping in flat spacetime, and then generalize the results to allow for cosmological expansion. In the first regime, maximal collisional damping of a gravitational wave, independent of the details of the collisions in the matter is, as we show, significant only when its wavelength is comparable to the size of the horizon.  Thus damping by intergalactic or interstellar matter for all but primordial gravitational radiation can be neglected. Although collisions in matter lead to a shear viscosity, they also act to erase anisotropic stresses, thus suppressing the damping of gravitational waves. Damping of primordial gravitational waves remains possible. We generalize Weinberg's calculation of gravitational wave damping, now including collisions and particles of finite mass, and interpret the collisionless limit in terms of Landau damping.  While Landau damping of gravitational waves cannot occur in flat spacetime, the expansion of the universe allows such damping by spreading the frequency of a gravitational wave of given wavevector.
    
\end{abstract}

\maketitle

\section{Introduction}
The opening of a new window on the universe through the ongoing observations of gravitational waves \cite{gwdetect} underlines the importance of reexamining
how they propagate through the matter in the universe, and  asking
what gravitational wave measurements can teach one about this matter.    Half a century ago, Hawking showed that if matter could be treated in the hydrodynamic limit the damping rate of a gravitational wave would be $\gamma = 16\pi G\eta$, where $G$ is Newton's gravitational constant and $\eta$ the viscosity of the matter \cite{Hawking, WeinbergBook}.  Using this result, Goswami et al.~\cite{goswami} argued that
gravitational wave observations could be used to constrain the viscosity of dark matter between the source and Earth.    But, as Hawking first pointed out, there are in general too few collisions in matter for hydrodynamics to be valid, and 
the damping would be less than the hydrodynamic result. 
Reference~\cite{lightman} estimated damping in the almost collisionless limit  by investigating the response of individual particles to a gravitational wave and found that the damping rate of the wave by nonrelativistic particles is 
\beq
  \gamma \sim  \frac{G nm}{\omega^2}\left(\frac{\bar v}{c}\right)^2 \frac{1}{\tau}.
\eeq 
Here $\omega$ is the frequency of the wave, $n$ the particle density, $m$ the particle mass, $\bar v$ the typical particle velocity, and $\tau$ the particle-particle collision time; the damping is $\sim 1/(\omega \tau)^2$ smaller than the viscous result. 

In addition to damping by collisions in matter, gravitational waves can also be attenuated by Landau damping, in which particles surf the gravitational wave and extract its energy, first proposed for gravitational waves in Ref.~\cite{GayerKennel}.  This effect was originally investigated in the context of plasma physics \cite{AK}, then in galactic dynamics \cite{lyndenbell}, and later in quantum chromodynamic plasmas \cite{screen}.     In a static flat universe massive particles cannot produce Landau damping  since they move more slowly than a gravitational wave.  In an expanding universe, however, Landau damping becomes possible, as we show, since the expansion in the presence of matter effectively spreads the  frequency of a gravitational wave.  
    Indeed the damping of cosmological gravitational waves by non-interacting neutrinos, as first proposed by Weinberg \cite{weinberg} and expanded upon in Refs.~\cite{WatanabeKomatsu,StefanekRepko},  can in fact be understood in terms of Landau damping, as we indicate below.   
   
     Our aim in this paper is to present a unified treatment of the damping of gravitational waves by matter, for arbitrary collision rates, thus encompassing the hydrodynamic and nearly collisionless limits studied earlier, as well as cosmological expansion.  We begin, in Sec.~\ref{static}, by considering a weak gravitational wave propagating through a dilute gas of colliding particles of arbitrary mass in an otherwise flat spacetime, and calculate, in Sec.~\ref{boltzmann}, the response of the matter to the wave using the Boltzmann equation.  For simplicity we work in the relaxation or collision time approximation. 
     
      As we show, in Sec.~\ref{max}, the maximum damping of the amplitude of a gravitational wave with frequency $\omega$ is or order $1/(\omega\tau_U)$ where $\tau_U$ is the age of the universe. Thus collisional damping by matter of gravitational waves generated by astrophysical sources cannot provide useful information about the nature of matter in the universe.  Furthermore, damping by dense environments surrounded localized sources of gravitational radiation is, as we estimate, insignificant.  After a general discussion of Landau damping in Sec.~\ref{landau},
we generalize the Boltzmann equation results in Sec.~\ref{cosmo} to describe collisional damping by particles of arbitrary mass in the presence of an expanding cosmological background. 
 
\section{Static spacetime \label{static}}

    Initially, we do not include the expansion of the universe, and consider rather the Minkowski space metric with a gravitational wave superimposed:
\beq
  ds^2  = - dt^2 + g_{ij}dx^idx^j,
  \label{flat}
\eeq
where
\beq
g_{ij} = \delta_{ij} + h_{ij}(\vec r, t),
\eeq 
with $h_{ij}$ the weak metric perturbation caused by a gravitational wave.   We work with $h_{ij}$ in the transverse--traceless gauge, and generally set $c=1$. 

   The effects on a gravitational wave passing through matter are given in terms of the gravitational wave equation in the transverse traceless gauge
\beq
    \partial_\mu \partial^\mu h_{ij} = \left(-\frac{\partial^2}{\partial t^2}+ \nabla^2 \right) h_{ij} = -16\pi G \pi_{ij},
\label{gweq}
\eeq    
where $\pi_{ij}$ is the transverse traceless part of the matter stress tensor,   $T_{ij,M}$, defined by
\beq
  \pi^i_{j}  \equiv  T^i_{j,M} - \frac{\delta^i_j }3 \sum_{k=1}^3 T^k_{k,M}.
  \label{ttt}
\eeq
In equilibrium,  $\frac13 \sum_{k=1}^3 T^k_{k,M}$ is simply the pressure $P$ of the matter.  

   The effect of a gravitational wave on a particle is given in terms of the dispersion relation,
\beq  
   p_\mu p_\nu g^{\mu\nu} + m^2 = 0,
\eeq 
which in the present case implies that the particle energy $\epsilon$ is given by
\beq
    \epsilon^2 = g^{ij}p_i p_j + m^2.
    \label{epsilon}
\eeq
Thus a weak gravitational wave changes the particle dispersion relation from $\epsilon_0 = \sqrt{p^2+m^2}$  to
\beq
    \epsilon=  \epsilon_0 + \delta\epsilon.
\eeq 
To first order in $h_{ij}$
\beq
\delta\epsilon = \frac12h^{ij}\frac{p_ip_j}{\epsilon_0} = - \frac12 h_{ij}\frac{p_ip_j}{\epsilon_0},
\label{deltae}
\eeq
since to this order, $h^{ij} = -h_{ij}$.  \\ \\

\section{Boltzmann equation \label{boltzmann}}

   We treat the matter as a dilute gas and calculate  $\pi _{ij}$ from the Boltzmann equation for the matter.
We first write the non-linear Boltzmann equation for the particle distribution function $f(r^i, p_j)$ as a function of the particle positions and canonical momenta,
\beq
  \left(\frac{\partial}{\partial t} + \vec \nabla_p \epsilon \cdot \vec \nabla_{r} -\vec \nabla_r \epsilon \cdot \vec \nabla_p\right) \,f  = {\cal C}, 
  \label{benonlin}
\eeq
where $\cal C$ is the collision term.  Here position gradients are taken with respect to $r^i$, and momentum gradients with respect to $p_i$.  This form of the equation is valid for relativistic as well as non-relativistic particles.   

The conservation laws of energy and momentum are found by taking the moments of Eq.~(\ref{benonlin}) with respect to $p_i$ and $\epsilon$; assuming that collisions conserve the total energy and momentum of the particles, we find (as in standard Fermi liquid theory)
\beq
 \frac{\partial}{\partial t} \int_p \epsilon f + \nabla_i \int_p \epsilon v^i f = \int_p \frac{\partial \epsilon}{\partial t} f,
 \label{E}
\eeq
and
\beq
 \frac{\partial}{\partial t} \int_p p_i f + \nabla_j T^j_{i,M} = - \int_p (\nabla_{r_i} \epsilon) f,
\label{p}
\eeq
where $\int_p \equiv g\int d^3p/(2\pi)^3$, with $g$ the number of internal states, e.g., spin, and
\beq
   T_{i,M}^j =   \int_p p_i v^jf =g^{jk} \int_p \frac{p_i p_k}{\epsilon}f 
   \label{TM}
   \eeq
is the matter stress tensor.  

   With Eq.~(\ref{deltae})  the right side of Eq.~(\ref{E}) becomes
\beq
   \int_p \frac{\partial \epsilon}{\partial t} f = \frac12\frac{\partial h^{ij}}{\partial  t} \int _p\frac{p_i p_j}{\epsilon}f = \frac12\frac{\partial h^{ij}}{\partial  t} T_{ij,M},
   \label{econs}
\eeq
so that the change in energy of the matter is given by
\beq
   \frac{\partial E}{\partial t} = \frac12\int d^3r\frac{\partial h^{ij}}{\partial  t} \pi_{ij}.
   \label{dedt}
\eeq
Only the transverse-traceless part of the stress tensor, Eq.~(\ref{ttt}), enters Eq.~(\ref{econs}).

Note that, for a wave of the form $H(z-ct)$,  say,  the right side of Eq.~(\ref{p}) becomes 
\beq
   -\int_p \frac{\partial \epsilon}{\partial z} f = \frac{1}{2}\frac{\partial h^{ij}}{\partial  t}  \pi_{ij},
   \label{momcons}
\eeq
indicating that as momentum $q$ is transferred from the gravitational wave, energy $q$ is also transferred.   Since $\pi^{ij}$ is
itself at least of first order in $h_{ij}$ the energy and momentum transfers are second order and higher in the amplitude of the gravitational wave. 
  
   We turn now to calculating the transverse-traceless part of the matter stress tensor, Eq.~(\ref{TM}); to linear order in $h_{ij}$,  
    \beq
  \delta T_{ij,M}  =  \int_p \frac{p_i p_j}{\epsilon_0}\left[\delta f - \frac{\delta\epsilon}{\epsilon_0}f_0\right] 
  \label{tijm}
     \eeq   
where $\delta f = f-f_0$, with $f_0$ the distribution function in the absence of $h_{ij}$.  The $\delta\epsilon$ term arises from the 
dependence of $\epsilon$ in the denominator on $h_{ij}$, Eq.~(\ref{deltae}).    Subtracting out the trace, we find, after using the vanishing trace of $h_{ij}$ and writing the equilibrium pressure of the matter as $\int_p (p^2/3\epsilon_0)f_0$, that 
\beq
    \pi_{ij}   =  \int_p \frac{p_i p_j}{\epsilon_0}\left[\delta f - \frac{\delta\epsilon}{\epsilon_0}f_0\right] +h^{ij}\int_p \frac{p^2}{3\epsilon_0}f_0. 
    \label{noname}
\eeq
The second term of this expression is manifestly traceless;  the trace of the first term vanishes since the integrations over  both $\delta f$ and $\delta \epsilon$, being symmetric in angles, vanish.  

   The latter two terms that contain $f_0$ can be simply combined, with an integration by parts using the transverse-traceless structure of $h_{ij}$, into a term proportional to $\partial f_0/\partial \epsilon$ so that Eq.~(\ref{noname}) becomes, 
 \beq
    \pi_{ij}   =  \int_p \frac{p_i p_j}{\epsilon_0}\left[\delta f - \delta\epsilon\frac{\partial f_0}{\partial \epsilon}\right] . 
    \label{noname1}
\eeq
This combination of terms falls out naturally, as we shall see, from the Boltzmann equation.

   
   Collisions between the particles, prior to freeze-out, tend to bring the distribution function into a local equilibrium in the presence of $h_{ij}$: 
\beq
     f \to f_h = \frac{1}{e^{\beta(\epsilon - \mu)}\mp 1},
     \label{hh}
 \eeq
 where $\epsilon$, given by Eq.~(\ref{epsilon}), depends on $h_{ij}$;  $\beta$ is the inverse temperature, and 
 $\mu$ the particle chemical potential.  Note that to first order in $h_{ij}$,
\beq
   f_h = f_0 +\delta\epsilon\frac{\partial f}{\partial \epsilon}.
\eeq
 For simplicity we employ a collision time approximation.  Since the only disturbances relevant here involve spherical harmonics of degree greater than one, we can write the collision term as  
\beq
  {\cal C} = - \frac{f-f_h}{\tau} = -\frac{1}{\tau}\left(\delta f -\delta\epsilon\frac{\partial f}{\partial \epsilon}\right),
\eeq 
where $\tau$ is the collision time, and $\delta f -\delta\epsilon\,\partial f/\partial \epsilon$ is the deviation of the distribution from local equilibrium;
 the additional terms commonly introduced to ensure conservation of particle number and total momentum (which involve spherical harmonics of degree zero and one) are not relevant \cite{AK}.  The linearized Boltzmann equation then reduces to
\beq
  \left(\frac{\partial}{\partial t} +\frac{1}{\tau}+\vec v \, \cdot \vec \nabla_{\vec r} \right) \,\delta f  &=& \left(\vec v\cdot\nabla_r \delta\epsilon + \frac{1}{\tau}\delta \epsilon\right) \frac{\partial f_0}{\partial \epsilon}.\nonumber\\
  \label{be3}
\eeq
With $h_{ij}(\vec r,t) = e^{i(\vec q\,\cdot \vec r-\omega t)}h_{ij}$, and  Fourier transforming in space and time we find
the solution of Eq.~(\ref{be3}),
\beq
  \delta f =  \frac{\partial f_0}{\partial \epsilon}
 \left( \frac{ -\vec q\cdot \vec v +i/\tau}{\omega - \vec q\cdot \vec v +i/\tau}\right) \delta\epsilon .
\eeq


  The deviation from local equilibrium is thus given by
\beq
  \delta f - \frac{\partial f_0}{\partial \epsilon}\delta \epsilon
    = -  \left( \frac{ \omega}{\omega - \vec q\cdot \vec v +i/\tau}\right)\frac{\partial f_0}{\partial \epsilon} \delta\epsilon .
\eeq
We then find the general result,
\beq
     \pi_{ij}(q,\omega)  =   \int_p \frac{p_i p_j}{\epsilon_0} \delta\epsilon \left(\frac{\omega}{\vec q\cdot \vec v -\omega -i/\tau}\right)\frac{\partial f_0}{\partial \epsilon}.\nonumber\\
    \label{deltatxy1}
   \eeq
Fourier transformed back to time, the stress tensor is 
\beq
     \pi_{ij}(q,t)  &&=  \\
     &&  -\int_p \frac{p_i p_j}{\epsilon_0} \int_{-\infty}^t dt' e^{-(iq\cdot v + 1/\tau)(t-t')}\frac{\partial f_0}{\partial \epsilon}\dot{\delta\epsilon}(t') .\nonumber
    \label{deltatxy2}
\eeq   
 
The response can be written in the more general form
\beq
      \pi_{ij}(q,\omega)&=& -\omega h_{ij} \int \frac{d\omega'}{2\pi} \frac{A(q,\omega')}{\omega-\omega' + i\xi}   ,
      \label{piA}
\eeq
where the spectral function is
\beq
A(q,\omega)  = -2 \int_p  \left(\frac{p_x p_y}{\epsilon_0}\right)^2  \frac{1/\tau} {(\omega -\vec q\cdot \vec v)^2+1/\tau^2 }\frac{\partial f_0}{\partial \epsilon}, \nonumber\\
\label{A}
\eeq
and $\xi$ is a positive infinitesimal.   In the collisionless limit, $1/\tau \to 0$, and for relativistic particles,
\beq
  A(q,\omega) = \pi\rho \langle (1-\zeta^2)^2 \delta(\omega - q\zeta)\rangle,
  \label{a}
\eeq
where $\rho$ is the energy density of the excitations, and the angular brackets denote the average over $\zeta \equiv \hat q \cdot \hat v$.    Fourier transformed back to time, 
\beq
    \pi_{ij}(q,t)&=& -\int_{-\infty}^t dt' \int \frac{d\omega}{2\pi} e^{-i\omega(t-t')}A(q,\omega) \dot h_{ij} (t'). \nonumber\\
 \label{pit}
\eeq     
     
     The damping of gravitational waves is governed by the imaginary part of the response,
\beq
  \Im\left(\frac{ \pi_{ij}}{h_{ij}}\right)  =
  - \omega \int_p  \left(\frac{p_i p_j}{\epsilon}\right)^2  \frac{1/\tau}{(\omega -\vec q\cdot \vec v)^2+1/\tau^2 }\frac{\partial f_0}{\partial \epsilon}. 
\label{deltatxy3}
     \eeq

   In the collision-dominated regime,  $\tau \ll 1/\omega$; doing the angular averages in the integrals we have
\beq
    \pi_{ij}  & = &  i\tau\omega \int_p  \frac{p_i p_j}{\epsilon_0} \delta\epsilon \,\,\frac{\partial f_0}{\partial \epsilon}
   = -\frac{ i\tau\omega}{15} \int_p  \frac{p^4}{\epsilon_0^2} \frac{\partial f_0}{\partial \epsilon} h_{ij}  \nonumber\\
   &&= -\eta \dot h_{ij}.
    \label{colldom}
\eeq
The viscosity calculated in the relaxation time approximation is
\beq
\eta =-\int_p  \left(\frac{p_i p_j}{\epsilon_0} \right)^2 \,\,\frac{\partial f_0}{\partial \epsilon_0}\tau,
\eeq
with $i\ne j$.    In this limit the damping rate of a gravitational wave, from Eq.~(\ref{gweq}), is
 $16\pi G \eta$, in agreement with earlier hydrodynamic treatments~\cite{Hawking, WeinbergBook}.

  For non-relativistic matter, $T\ll mc^2$, the $\vec q\cdot \vec v$ in the denominator  of $A$ can be neglected,  and we have
\beq
    \pi_{ij}  \simeq   \frac{\omega}{\omega+i/\tau}P\,h_{ij}.
   \label{deltaxynonrel}
  \eeq

 From the imaginary part of Eq.~(\ref{gweq}),  the dispersion relation of gravitational waves 
 is
 \beq
   \omega \simeq q + \frac{8\pi G}{\omega}\frac{\pi_{ij}}{h_{ij}},
\eeq
so that the damping of a wave is given by
\beq 
 \Im\omega = \frac{8\pi G}{\omega}\Im\left(\frac{\pi_{ij}}{h_{ij}}\right).
 \label{damping}
\eeq 
For fully relativistic matter, in the nearly collisionless limit, to first order in $1/\tau\omega$, 
  \beq
       \pi_{ij}  =  -P\left(1-\frac{2i}{\tau\omega}\right) h_{ij},
    \label{deltatxyrel}
\eeq
while in the collision-dominated regime, $\pi_{ij}$ is given by Eq.~(\ref{colldom}).

\section{Maximum collisional damping \label{max}}

As one can see from Eq.\ (\ref{deltatxy3}), $\Im(\pi_{xy}/h_{xy})$ has its maximum magnitude for $\omega\tau \sim 1$.  The possible damping is thus limited by
 \beq
 \label{mcd}
  {\rm Max}(|\Im \omega|) \lesssim  -\frac{8\pi G}{\omega} \int_p \frac{p^4}{15\epsilon_0^2} \frac{\partial f_0}{\partial \epsilon} 
  \le \frac{8\pi GP}{\omega},
\eeq 
where $P$ is the total local pressure of the matter under consideration, which gives rise to the damping.  It is instructive to write this bound in terms of the 
age of the universe, defined by 
\beq
    \frac{1}{\tau_U^2} = \frac{8\pi G}{3}
      \bar\rho =   \left(\frac {\dot a}{a}\right)^2, 
\eeq   
where $a$ is the cosmological scale parameter and $\bar\rho$ the mean mass density of the universe.  Since the
mean pressure obeys $\bar P\le \bar\rho/3$, we find
\beq
     {\rm Max}(|\Im \omega|)   \lesssim \frac{P}{\bar P}\frac{1}{\omega\tau_U^2}.
\eeq
A wave traversing matter will average the local pressure, and thus we can conclude,
\beq
     {\rm Max}(|\Im \omega|)   \lesssim  \frac{1}{\omega\tau_U^2},
     \label{maxdamp}
\eeq
indicating that damping of a gravitational wave by matter in the universe can only be significant for a wave of frequency of order
$1/\tau_U$.  This bound includes all contributions from dark matter particles as well \cite{goswami}.

 To express the result (\ref{maxdamp}) in another way, the collisional damping of a gravitational wave within the characteristic expansion time of the universe is of order $1/\omega \tau_U$.   For  $\omega \sim 10^3 $ s$^{-1}$, as in the recent gravitational wave detections \cite{gwdetect}, and $\tau_U \sim 10^{18}$ s, this ratio is $\sim 10^{-21}$.   Collisional damping in intergalactic or interstellar matter of gravitational waves produced by astrophysical sources  is not useful to determine the nature of matter in the universe.   This result is valid for any particle-like form of dark matter, including that in a possible shadow universe \cite{nishijima} or matter that only interacts with gravitationally suppressed interactions \cite{sloth}.  Furthermore, collisional damping in locally high dense environments, e.g., in the neighborhood of mergers of black holes or neutron stars,  is also negligible for gravitational waves produced by astrophysical sources, as we argue in the next section.  On the other hand for primordial gravitational waves with $\omega \sim 10^{-16}-10^{-15}$, one has $1/\omega\tau_U \sim 10^{-3}-10^{-2}$, an effect that could play a role in interpretation of future precision measurements of the spectrum of primordial gravitational radiation.\footnote{We thank Vicky Kalogera and Chris Pankow for this observation.}  

As we discuss in the following sections, Landau damping cannot occur in a flat spacetime.  Even in an expanding universe,
the Landau damping rate is  $\sim 1/ \omega^2 t_U^3$, so that the total damping within the expansion time of the universe is
$\sim 1/(\omega t_U)^2$, a factor $1/\omega t_U$ smaller than the maximum collision damping.

One can write the contribution to the damping from a particular component, $s$, of the matter, e.g., neutrinos or dark matter,
in the form 
\beq
  |\Im \omega|_s \equiv n_s\sigma_{GW,s},
\eeq
where $\sigma_{GW,s}$ is the graviton scattering cross section on particles of species $s$, and $n_s$ is the number density of
species $s$.  The ratio of the cross section to the Planck length, $\ell_{Pl}$, squared is essentially bounded above by
\beq
   \frac{\sigma_{GW,s}}{\ell_{Pl}^2}   \lesssim  \frac{\langle pv\rangle_s}{\omega} 
\eeq
where $\langle pv\rangle_s$ is the mean product of the particle momentum and velocity of species $s$, which is of order the
temperature for a species in thermal equilibrium, or the temperature at which the species froze out.   Thus in general, 
\beq
     \frac{\sigma_{GW,i}}{\ell_{Pl}^2}   \lesssim \frac{T}{\omega},
\eeq
with the above understanding of $T$.  

\section{Maximal collisional damping in dense environments  \label{dense}}

We look now at the damping of gravitational wave produced by binary astrophysical sources, as the waves pass through the dense medium surrounding the sources.   Collisional damping is limited by $|\Im \omega| < \gamma_{\rm max}$, where from Eq.~(\ref{mcd}),
\beq
\gamma_{\rm max} \sim \frac{1}{\mpl^2}\frac{P}{\omega} = \frac{w}{\mpl^2}\frac{\rho}{\omega};
\eeq    
we have introduced the Planck mass $\mpl^{-2} = 8\pi G$ and written the relation between the pressure and the energy density  by the equation of state parameter $w = P/\rho$. 

Assuming the gravitational wave source to a binary system inside a region surrounded by matter with a given density profile with equation of state parameter $w$, we find collisional damping along a line of sight to be significant if
\eq{}{\int_0^R\, dr\, \gamma_{\rm max} \sim 1}
where $R$ is a physical radius enclosing the ambient matter.    A reasonable first estimate is simply to associate the integral with the characteristic size $R_c$ of the dense region:
\beq
\int_0^R\, dr\, \gamma_{\rm max}  \sim \frac{w}{\mpl^2}\frac{\rho}{\omega}R_c \sim  \frac{w}{\mpl^2R_c^2}\frac{M}{\omega}  
\eeq
where $M$ is the total mass in the region with characteristic size $R_c$. To go beyond this estimate, one could take the density profile from detailed calculations, e.g. the profile of a typical dark matter halo (determined via phenomenological models that fit N-body simulations such as the Navarro-Frenk-White or Einasto density profiles \cite{halo}), and find numerical factors that little affect the  conclusion.  In ``natural" units, one solar mass $M_\odot \sim 10^{66}$ eV, 1 Hz $\sim 10^{-15}$ eV, $\mpl^{-1} = \ell_{Pl} \sim 10^{-35}$ m, and 1 kpc $~\sim 10^{19}$ m, one has 
\beq
\int_0^R\, dr\, \gamma_{\rm max}  \sim w \frac{10^{-27}}{\left(R_c/{\rm kpc}\right)^2}\left(\frac{M}{M_\odot}\right)\frac{1}{\left(\nu/{\rm Hz}\right)},
\eeq
where $\nu = \omega/2\pi$. We first consider a typical galactic halo surrounding a binary system source of the gravitational wave. Here, typically $M \sim 10^{12}M_\odot$,  $R_c \sim 100$ kpc, so that 
\beq
\int_0^R\, dr\, \gamma_{\rm max}  \sim w \frac{10^{-19}}{\left(\nu/{\rm Hz}\right)} 
\eeq   
which is feeble for all astrophysical sources within the halo.

We next consider the ambient region surrounding a binary system similar to that which gave rise to GW150914 -- containing a dense distribution of  (not necessarily dark) matter.  If the ambient matter has a mass comparable to that of the binary system localized within some region $R_c \gg R_s$,  the Schwarzschild radius associated with the total mass of the binary system, we find that at the lowest frequencies of the binary system, the factor $M/M_\odot \times \left(\nu/{\rm Hz}\right)^{-1}$ is $\mathcal O(1)$; furthermore for the expected non-relativistic low pressure surrounding matter, $w \sim c_s^2$, the square of the adiabatic sound velocity.  Thus
\beq
\int_0^R\, dr\, \gamma_{\rm max}  \sim c_s^2 \frac{10^{11}}{\left(R_c/{\rm m}\right)^2}.
\eeq
which can only be of order unity if the ambient matter is localized to within a radius $R_c \lesssim c_s\times 300\,{\rm km}$ around the binary system,
which even for mildly non-relativistic ambient matter (e.g. $c_s \sim \mathcal O(10^{-1})$) would require a mass comparable to that of the binary system to be crammed into a region comparable to the Schwarzschild radius of the final merged black hole ($\sim 70$ km $\sim R_s$); such a high density is contrary to our initial assumption that $R_c \gg R_s$.  Requiring that this matter be distributed within a region an order of magnitude larger than the binary system yields $\int_0^R\, dr\, \gamma_{\rm max}  \sim c_s^2 \ll 1$.   We conclude that a distribution of non-relativistic matter of high density surrounding the source of gravitational radiation is not capable of significantly damping gravitational radiation. 

More realistically, one is in general far from the condition of maximal collisional damping, that the collision rate $\tau^{-1}$, be comparable to the frequency of the gravitational wave.   Maximal collisional damping is a highly unlikely prospect even for relativistic matter jets and lobes  close to the merger of neutron star/black hole binary systems  \cite{stu1,stu2}. 
 To see this we write roughly, $ \tau^{-1} =  n \sigma v$, where $n$ is the density of particles, $\sigma$ is a particle-particle scattering cross section, and $v$ a mean particle velocity. In terms of the mass, $M$, and characteristic radius, $R_c$, of the dense environment,
\beq
\frac{1}{\tau} \sim  \frac{M}{M_\odot}\frac{10^{35}}{\left(R_c/{\rm m}\right)^3} \left(\frac{\sigma}{{\rm fm}^2}\right)\frac{v}{c}\, {\rm s^{-1}},
\eeq
 Clearly, for a typical nuclear or particle physics cross section, the above is much larger than the typical frequency of an astrophysical binary system by many orders of magnitude, so that $\omega \tau \ll 1$.

\section{Landau damping: general considerations  \label{landau}}

In flat space in the collisionless limit ($\tau \to\infty$)
 Eq.\ (\ref{deltatxy3}) reduces to 
\beq
     \Im\left(\frac{ \pi_{ij}}{h_{ij}}\right)  =\pi\omega \int _p \left(\frac{p_i p_j}{\epsilon_0}\right)^2  \delta(\omega -\vec q\cdot \vec v\,)\frac{\partial f_0}{\partial \epsilon},
\label{deltatxy4}
     \eeq
a result describing Landau damping,  the decay of the mode into a single particle--hole pair.\footnote{In  the language of quantum mechanics, the damping may be regarded as the creation, with amplitude $\propto 1/(\omega -\vec q\cdot \vec v\,s)$, of a virtual single particle--hole pair, which subsequently decays  into two real particle--hole pairs.  Equation~(\ref{deltatxy3}) can be understood, when $1/\tau \to 0$, as this amplitude squared, summed over all momenta.} The particle-hole excitations are spacelike.
For a gravitational wave, $\omega=q$, the integral vanishes except possibly for massless particles moving in the same direction, say $\hat z$, as the gravitational wave.  However, for such particles, the factor $p_i^2p_j^2 \to p_x^2p_y^2$ vanishes; Landau damping is forbidden in the absence of cosmological expansion.  Following general remarks on Landau damping in this section we show in the following section how the collisionless damping process described by Weinberg \cite{weinberg}  can be understood as a generalization of Landau damping, driven by the expansion of the universe.

     In an expanding universe, the gravitational wave energy changes during expansion, i.e., the frequency of the gravitational wave is not constant, since the expansion absorbs energy from the wave. This energy loss is different from Landau damping by the matter traversed by the wave.   When the phase velocity of the wave is different from the group velocity of the excitations in the matter, energy in flat spacetime is pumped to and fro between the wave and the matter, but the net rate of transfer is zero because the energy transferred in one half-cycle of the wave is exactly cancelled by the loss in the other half-cycle.  In an expanding universe, however, the cancellation is incomplete.   
     
    We recall the energy loss caused by expansion.  A weak gravitational wave of
period small compared with the age of the universe behaves in the absence of matter as
$h_{ij}  = \chi(u) e^{-iqu}$,  where $u$ is conformal time, related to coordinate time by  $du = dt/a(t)$;
as we see in the next section, $\chi(u) \propto 1/a(u)$.  This structure is expected on the basis of simple arguments:  the energy of a gravitational wave packet is proportional to the energy density in the wave packet times the volume of the packet.  The energy density of the wave varies as $g^{ii}(\partial \chi/\partial x^i)^2 \sim a^{-4}$  and the volume of the packet varies as $a^3$, so the total energy decreases as $1/a$. This result also agrees with simple redshift arguments:  The energy of a massless particle varies as $1+z \sim 1/a$  owing to the expansion of the universe, and thus the energy density measured in locally Minkowskian spacetime ($ds^2=-dt^2+(d{\vec r}\,)^2$) varies as $1/(1+z)^4$.   For example, the redshift of the first LIGO event GW150914 was $z=0.09^{+0.03}_{-0.04}$ \cite{gwdetect}, leading to an energy density reduction by a factor $\simeq 1-1/(1.09)^4 \approx 30\%$ from cosmological expansion.   By comparison,  even were the intervening matter collisionless, Landau damping of the wave would be totally negligible, $\sim 1/(\omega t_U)^2$.

\section{Gravitational wave damping with cosmological expansion \label{cosmo}}

     We turn now to relate the gravitational radiation damping derived by Weinberg \cite{weinberg} to the calculations above, and to Landau damping in collisionless plasmas, driven by the expansion of the universe.   We first generalize the treatment of \cite{weinberg} to allow for massive particles and collisions, working in conformal time, $u$, related to coordinate time by
$dt = a\,du$, where $a$ is the scale parameter of the expansion.  The metric in the presence of expansion and a gravity wave is given by
\beq
    ds^2 = a(u)^2[-du^2 + (\delta_{ij} + h_{ij})dx^idx^j].
    \label{ga}
\eeq
Owing to the explicit $a^2$, upper and lower components of vectors are related by $x_\mu= a^2 x^\mu$ to zeroth order in $h$.
In addition, the energy of a particle in the metric (\ref{ga}) is given,  in the absence of $h_{ij}$, by
\beq 
      \epsilon_0^2 = p_i^2+ a^2m^2.
\eeq
We study, following \cite{weinberg}, the evolution of the coupled gravitational wave -- matter system, after an initial time $u_0$ at which 
the matter distribution function is given as $f_0$, essentially the $f_h$ in Eq. (\ref{hh}).   Since $f_0$ includes the metric perturbations
$h_{ij}(u_0)$ at that time, the additional perturbations of the energy that modify the distribution function by $\delta f(u)$ at later time depend only on the deviation from $h_{ij}(u_0)$, that is,
\beq
\delta \epsilon &=&  -\frac{p_ip_j}{2\epsilon_0}\left[h_{ij}(u) - h_{ij}(u_0)\right].
\eeq

   The Boltzmann equation in conformal time, Fourier transformed in space (cf. Eq.~(\ref{be3})) is
\beq
\left(\frac{\partial}{\partial u} + \frac{1}{\tau_c} + i\vec q\cdot\vec v \right)\delta f = \frac{\partial f_0}{\partial \epsilon}\left(\frac{1}{\tau_c} + i\vec q\cdot\vec v \right)\delta \epsilon,
\label{boltz}
\eeq
The particle velocity, $v$, the distribution function $f_0$, and the conformal collision time $\tau_c =\tau/a$, are directly dependent on the scale factor.  (More generally $\tau$ will depend on the cosmological epoch and thus contain further dependence on the scale factor, a question we do not pursue here.) In particular
\beq
  v^i= \partial \epsilon_0/\partial p_i = \frac{p_i}{\sqrt{p_jp_j + m^2 a(u)^2}}.
\eeq  
Similarly, an equilibrium distribution function,
\beq
    f_0 = \frac{1}{e^{\epsilon_0/T_0(u)} \mp 1}
\eeq
(with $\mp$ for bosons or fermions) depends on $a$ through the term $m^2a^2$ in $\epsilon$, and $T_0(u)/a(u)$ is the temperature of the dark matter.  For massless particles, $T_0$ is constant.   In addition $\tau_c$ is a function of the ambient density along the trajectory of the gravitational wave, and so in general depends on time through its dependence on the scale factor, as well as through the evolving particle distributions.   Equation~(\ref{boltz}) has the general solution
\beq
\delta f(u,p)&=& \int^u_{u_0}\,du' \frac{\partial e^{-\Phi(u,u')} }{\partial u'}
\frac{\partial f_0}{\partial \epsilon}(u') \delta\epsilon(u');
\label{bsol}
\eeq
where we write 
\beq
\Phi (u,u')&\equiv & \int^{u}_{u'} du''\,\left(\frac{1}{\tau_c(u'')} + i\vec q\cdot\vec v(u'')\right)\nn\\
&& = \ell(u,u') + i\hat q\cdot\hat p \,s(u,u'),
\eeq 
in terms of 
\beq
\ell(u,u') = \int^{u}_{u'}\frac{du''}{\tau_c(u'')},
\eeq
and 
\beq
  s(u,u')= q\int_{u'}^u du'' v(u''),
\eeq
which is the displacement of the particle in the interval $u'$ to $u$ times the wavevector.
The deviation from local equilibrium in (\ref{pi}) is therefore
\beq
 \delta f - \frac{\partial f}{\partial \epsilon}\delta\epsilon& =&-\int^u_{u_0}\,du' e^{-\Phi (u,u')}
\frac{\partial}{\partial u'}\left[\frac{\partial f_0}{\partial \epsilon} \delta\epsilon(u') \right].  \nn\\
 \label{source}
\eeq

  On a cosmological background, 
\beq
\pi_{ij} = \int_p \frac{p_ip_j}{\sqrt{-g}\,\epsilon_0}\left[\delta f - \frac{\partial f}{\partial \epsilon}\delta\epsilon\right],
\label{pi}
\eeq
the generalization of Eq.~(\ref{noname1}), where in the absence of a gravitational wave, $\sqrt{-g} = a^4$.  With 
Eq.~(\ref{source}), we then have
\beq
   \pi_{ij}(q,u)& =&-\int_p \frac{p_ip_j}{a(u)^4\,\epsilon_0}
   \int^u_{u_0}\,du' e^{-\Phi (u,u')}
\nn\\ &&\hspace{12pt}\times\frac{\partial}{\partial u'}\left[\frac{\partial f_0}{\partial \epsilon} \delta\epsilon(u')\right].  
 \label{pi1}
\eeq
This equation is the direct generalization of Eq.~(\ref{deltatxy2}) to an expanding spacetime.

   Using $\epsilon d\epsilon = p dp$ we have
\beq
\pi_{ij} =  &&\hspace{-9pt}\frac{1}{a(u)^4}\int_p\frac{p_ip_jp_kp_l}{2p\epsilon_0(u)} \nn\int^u_{u_0}\,du' e^{-\Phi(u,u')}\\&& \times \frac{\partial}{\partial u'}\left[\frac{\partial f_0}{\partial p} \left(h_{kl}(u') - h_{kl}(u_0)\right)\right]. 
\eeq
Since $k\ne l$ the angular average above has the form $(\delta_{ik}\delta_{jl}+\delta_{il}\delta_{jk})K(s)$ where in terms of spherical Bessel functions $K(s) = j_2(s)/s^2$, and explicitly,
\beq 
K(s)\hspace{-5pt}&=&\hspace{-5pt}\int \frac{d\Omega}{4\pi} e^{-i\, \zeta\,s}\, (1-\zeta^2)^2 {\rm sin}^2\varphi\, {\rm cos}^2\varphi\nn\\ &=& \hspace{-5pt}-\frac{{\rm sin}\,s}{s^3} - 3\frac{{\rm cos}\,s}{s^4}  +  3\frac{{\rm sin}\,s}{s^5},
\eeq
with $\zeta=\cos \theta$;  thus
\begin{eqnarray}
\pi_{ij}(q,u)& =& \frac{1}{a(u)^4}\int_p\frac{p^3}{\epsilon_0(u)}\int^u_{u_0}\,du' e^{-\ell(u,u')} K(s)\nn\\
&&\times\frac{\partial}{\partial u'}\left[
\frac{\partial f_0}{\partial p}
\left(h_{ij}(u')-h_{ij}(u_0)\right)\right].
\label{genexp} 
\end{eqnarray}
In the massless limit, we integrate the momentum derivative by parts, using $\epsilon d\epsilon = p dp$, and noting that $s\to q(u-u')$, to obtain  
\beq
\pi_{ij} = -4\bar\rho \int^u_{u_0}\,du' e^{-\ell(u,u')} K\left(q(u-u')\right)  h_{ij}'(u'),
\label{massexp}
\eeq
where the prime denotes $d/du$, and 
\beq
 \bar\rho = \frac{1}{a^4} \int_p pf_0
\eeq
is the mass density of the matter. 
 Away from the massless limit, generalizing Ref.~\cite{weinberg}, we find extra contributions from the $p$ dependence of $s$ in $K$.    We see from Eq. (\ref{genexp}) or (\ref{massexp}), that, as expected, the net effect of collisional interactions is to efficiently erase anisotropic stresses, and hence limit their ability to damp gravitational waves.  
 
    Since astrophysical sources of gravitational waves have characteristic frequencies much greater than the inverse Hubble scale at late times, we can expand the time dependence of the mode functions in powers of $a'/(a q)$.   For such a wave, the spatial Fourier component $\vec q$ obeys the equation of motion,
 \beq
   h_{ij}''+2\frac{a'}{a} h_{ij}'+ q^2 h_{ij} = 16\pi G a^2  \pi_{ij}.
\label{gweqa}
\eeq
The solution for $h_{ij}(q,u)$ in the absence of matter is, to lowest order in $a'/(a q)$,
\beq
   h_{ij}(q,u) \propto \frac{e^{-iqu}}{a(u)}
   \label{h1}
\eeq
(during radiation domination, this result is exact).
We assume that the gravitational wave is in the form of a wavepacket for which 
\beq
  h_{ij}(\vec r,u) = \int \frac{d^3q}{(2\pi)^3} \frac{e^{-iqu}}{a(u)} F(q),
  \label{pack}
\eeq
where $F(q)$ is localized about a wave vector $\vec q_0$.   We consider the absorption by a region of matter much smaller than the horizon size.

   To see the mechanism of Landau damping in the absence of collisions, we calculate the damping of the wave directly in terms of the energy transfer to the matter, writing, from Eq.~(\ref{dedt}), using conformal time 
\beq
   \frac{\partial E}{\partial u} = - \frac12 \int d^3 r h'_{ij}(\vec r,u)\pi_{ij}(\vec r,u) \nonumber\\
   = - \frac12 \Re \int \frac{d^3q}{(2\pi)^3}  {h'_{ij}}^*(q,u)\pi_{ij}(q,u). 
   \label{etransf} 
\eeq
We work in the massless limit, in order to illustrate the physics with the fewest complications.  Then $f_0$ does not depend on $a$, and $\Phi(u,u') \to iq\zeta (u-u')$, and one has, 
\beq
   \frac{\partial E}{\partial u} & =& -\frac{1}{4a(u)^4 } \Re \int \frac{d^3q}{(2\pi)^3}  {h'_{ij}}^*(q,u)
     \int_p \frac{p_ip_jp_kp_l}{p^2}   \nn\\ &&\times
   \int^u_{u_0}\,du' e^{-i\vec q\cdot\hat p (u-u')}\frac{\partial f_0}{\partial p} h'_{kl}(q,u')\nn\\
   &=&-\frac{1}{16a(u)^4 } \Re \int \frac{d^3q}{(2\pi)^3}  {h'_{ij}}^*(q,u)
     \int_p p^2 \frac{\partial f_0}{\partial p} \nn\\ &&\times (1-\zeta^2)^2 
   \int^u_{u_0}\,du' e^{-iq\zeta (u-u')} h'_{ij}(q,u').
      \label{etransf} 
\eeq

    From Eq.~(\ref{h1}), we see that $h_{ij}'(u) = -(iq+{\cal H}(u))h_{ij}(u)$, where ${\cal H} \equiv a'(u)/a(u)$.  The explicit ${\cal H}(u)$, which is small relative to the $q$ term and for an astrophysical gravitational wave produces only a small correction to the Landau damping, can be neglected.    Over the time span of a gravitational wavepacket transversing a given region of matter, the scale factor $a(u')$ in $h_{ij}(u')$ can be expanded as $a(u') = a(u) + a'(u)(u'-u) \simeq a(u)e^{{\cal H}(u'-u)}$.      Thus the $u'$ integral can be written as 
\beq
&&\frac{-iqe^{-iqu}F(q)}{a(u)}\int_{u_0}^u   du' \,  e^{(iq(1-\zeta) +{\cal H}(u))(u-u') } \nn\\
& \simeq &\frac{1}{{\cal H}+iq(1-\zeta)}\left(h_{ij}'(q,u)-e^{-iq\zeta(u-u_0)}h_{ij}'(q,u_0)  \right). \nonumber\\
\eeq
Since the characteristic frequencies are large compared with $1/(u-u_0)$ the phase factor in the final term will average to
zero inside the $q$ and $\zeta$ integrals in Eq.~(\ref{etransf}).  We find then
\beq
  \frac{\partial E}{\partial u} &=&  \frac{\bar \rho}{4}  \int \frac{d^3q}{(2\pi)^3}  |h'_{ij}(q,u)|^2 \nn\\ &&\times
    \int_{-1}^1 \frac{d\zeta}{2}    \frac{{\cal H} (1-\zeta^2)^2}{{\cal H}^2 +q^2(1-\zeta)^2 }. 
     \label{detot1}
 \eeq
We see here how expansion of the universe introduces a spread in frequencies $\sim\pm{\cal H}$ about $q$, thus allowing Landau damping; in the absence of expansion, ${\cal H} = 0$, and Landau damping vanishes.

   To lowest order in ${\cal H}/q$ the integral is simply 4/3, so that
\beq
  \frac{\partial E}{\partial u} = \frac{\bar\rho}{3}\frac{a'}{a} \int \frac{d^3q}{(2\pi)^3} \frac{ |h'_{ij}(q,u)|^2}{q^2}.
   \label{dedt5}
\eeq
The energy density of the gravitational wave is
\beq
  E_{gw} = \int \frac{d^3q}{(2\pi)^3}\frac{|h_{ij}'(q,t)|^2}{32\pi G} ,
\eeq
so that for a wavepacket centered about a frequency $\bar q$
\beq
  \frac{\partial E}{\partial u} = \frac{32\pi G\bar\rho}{3\bar q^2}{\cal H} E_{gw}.
   \label{dedt5}
\eeq
Finally we note that $8\pi G\bar\rho /3\sim (a'/a^2)^2$ and thus
\beq
  \frac{\partial E}{\partial u} \sim \frac{4{\cal H}^3}{(a\bar q)^2}  E_{gw}.
   \label{dedt5}
\eeq
The characteristic absorption time via Landau damping is thus $\sim\omega^2 t_U^3$ (with $\omega=\bar q$), which is thoroughly negligible.  
The corresponding fractional change in energy over an expansion time of the universe is $\sim 1/(\omega t_U)^2$.

\section{Concluding remarks}

In this paper we have laid out a framework for evaluating the damping of gravitational radiation by matter with arbitrary mass particles and collision strengths. By considering the damping of gravitational waves in both flat spacetime and in an expanding universe, we identify two distinct mechanisms through with damping can occur -- the first in which collisions produce the damping, and the second, via Landau damping. 

   When the expansion of spacetime can be neglected, the damping of a wave of a given frequency, proportional to the {\it relaxation rate} $1/\tau$ in the collisionless regime ($\omega \tau \gg 1$) and to the {\it collision time}, $\tau$, in the hydrodynamic regime ($\omega \tau \gg 1$), is maximal when $\omega \tau \simeq 1$.  For the frequencies to which LIGO is sensitive and for plausible models of dark matter, calculations of damping based on hydrodynamical considerations are gross overestimates, and we conclude that it is impossible from current observations of gravitational waves to put useful bounds on the properties of dark matter.  Landau damping in this case is not possible because particles have velocities less than $c$.   As we estimate in Sec.~\ref{dense}, collisional damping of gravitational waves of frequencies produced by astrophysical binary systems, propagating through dense local environments, is also insignificant. 
     
Collisionless damping is possible in an expanding universe since the frequency of the gravitational wave and the energies of particles depend on time.  Damping of gravitational waves by free-streaming relativistic particles, proposed by Weinberg \cite{weinberg}, may, as we have shown, be regarded as a generalization of Landau damping; we have also generalized Weinberg's formalism to allow for collisions in the matter. We note in passing that one can straightforwardly extend the present framework to incorporate scenarios of non-thermal dark matter, since it was not essential to assume a specific functional form for the distribution, see, e.g., Eq.~(\ref{hh}).
   
In the future we will apply the present framework to study the processing of stochastic gravitational waves of primordial origin, e.g., from (first order) phase transitions in the early universe during matter domination, in scenarios of ultralight dark matter. Such scenarios are similar to damping by neutrinos during radiation domination \cite{weinberg, WatanabeKomatsu}, and might describe damping by axions \cite{hui,sikivie,marsh}.

\section*{Acknowledgments}  We are grateful to Stu Shapiro and Subir Sarkar for very helpful remarks.  The research of author GB was supported in part by NSF Grant PHY1305891 and PHY1714042.   He is grateful to the Aspen Center for Physics, supported in part by NSF Grants PHY1066292 and PHY1607611, and the Niels Bohr International Academy where parts of this research were carried out.  Author SP is supported by funds from Danmarks Grundforskningsfond under Grant No.~1041811001.


\begin{thebibliography}{20}


\bibitem{gwdetect}  B. P. Abbott et al. (LIGO Scientific Collab. and Virgo Collab.)
Phys. Rev. Lett. {\bf 116}, 061102 (2016); ibid. {\bf 116},  241103 (2016); arXiv:1706.01812.

\bibitem{Hawking} S. W. Hawking,  Astrophys. J. {\bf 145}, 544 (1966). 

\bibitem{WeinbergBook} S. Weinberg, {\em Gravitation and Cosmology: Principles and Applications of the General Theory of Relativity}, (Wiley, NY, 1972). Ch. X. 

\bibitem{goswami} G.~Goswami,  G. K. Chakravarty, S.~Mohanty, and A.~R.~Prasanna,  Phys. Rev. D {\bf 95}, 103509 (2017).

\bibitem{lightman}   A. P.  Lightman, W. H.  Press,  R. H. Price, and S. A. Teukolsky, {\em Problem Book in Relativity and Gravitation} (Princeton Univ. Press, 1975),
Problem 18.15.

\bibitem{GayerKennel}  S. Gayer and C. F. Kennel, Phys. Rev. D {\bf 19}, 1070 (1979).

\bibitem{AK} A. A. Abrikosov and I. M. Khalatnikov, Rep. Prog. Phys. {\bf 22}, 329 (1959), Eq. (10.1).

\bibitem{lyndenbell} D. Lynden-Bell, MNRAS {\bf 124}, 279 (1962).

\bibitem{screen}  G. Baym, H. Monien, C. J. Pethick and D. G. Ravenhall,  Phys.  Rev.  Letters 64, 1867 (1990).

\bibitem{weinberg}  S. Weinberg, Phys. Rev  D {\bf 69}, 023503  (2004).

\bibitem{WatanabeKomatsu}   T. Watanabe and E. Komatsu, Phys. Rev. D {\bf 73}, 123515 (2006).

\bibitem{StefanekRepko}  B. A. Stefanek and W. W. Repko, Phys. Rev. D {\bf 88}, 083536 (2013).

\bibitem{nishijima} K. Nishijima and M. H. Saffouri,  Phys. Rev. Lett. {\bf 14}, 205 (1965).

\bibitem{sloth} M. Garny, M. Sandora and M.~S.~Sloth, Phys. Rev. Lett.  {\bf 116}, 101302 (2016)

\bibitem{halo} A. W. Graham, D. Merritt, B. Moore, J. Diemand and B. Terzic, Astron. J.  {\bf 132}, 2685 (2006).

\bibitem{stu1} 
V. Paschalidis, M. Ruiz and S. L. Shapiro, 
Astrophys. J.  Letters {\bf 806}, L14:1-5, (2015).

\bibitem{stu2} M. Ruiz, R. Lang, V. Paschalids and S. L. Shapiro,
Astrophys. J. Letters {\bf 824}, L1:1-5 (2016).

\bibitem{hui}  L. Hui, J. P. Ostriker, S. Tremaine, and E. Witten, Phys. Rev. D {\bf 95}, 043541 (2017).

\bibitem{sikivie}  N. Banik and P. Sikivie,  in {\em Universal Themes of Bose-Einstein Condensation,} (eds.  D. Snoke, N. Proukakis and P. Littlewood, Cambridge Univ. Press, 2017).

\bibitem{marsh} D. J. E. Marsh, Phys. Repts. {\bf 643} 1-79  (2016).




\end{thebibliography}
\end{document}